\documentclass[]{relaxed_system_lab}

\usepackage[utf8]{inputenc}
\usepackage[T1]{fontenc}
\usepackage{charter}
\usepackage{varwidth}
\usepackage{wrapfig}
\usepackage{fontawesome5}

\usepackage{amsmath, amssymb, amsfonts, bm} 
\usepackage{graphicx}          

\usepackage{url}               
\usepackage{xspace}            
\usepackage{enumitem}          

\usepackage{makecell}          

\usepackage{algorithm}
\usepackage{algpseudocode}

\usepackage{float}
\usepackage{soul}
\usepackage{pgfplots}
\usepackage{pgfplotstable}

\usepackage{natbib}
\usepackage{latexsym}

\usepackage{url}
\usepackage{amssymb}
\usepackage[utf8]{inputenc}
\usepackage{microtype}
\usepackage{booktabs}
\usepackage{pifont} 
\usepackage{multirow}
\usepackage{makecell}
\usepackage{xspace}
\usepackage{color}
\usepackage{xcolor}
\usepackage{colortbl}
\usepackage{adjustbox}
\usepackage{hyperref} 
\usepackage[edges]{forest}
\usepackage{tikz} 
\usepackage{caption}
\usepackage{amsfonts}

\hypersetup{
    colorlinks,
    linkcolor={blue!80!black},
    citecolor={blue!80!black},
}
\tikzset{
    root/.style =             {align=center, text width=1cm, rounded corners=3pt, line width=0.3mm, fill=gray!10, draw=gray!80, font=\small},
    demographic/.style =         {align=center, text width=1.8cm, rounded corners=3pt, line width=0.3mm, fill=blue!10, draw=blue!80, font=\footnotesize},
    demographic_work/.style =    {align=center, text width=10cm, rounded corners=3pt, line width=0.3mm, fill=blue!10, draw=blue!0, font=\footnotesize},
    character/.style =         {align=center, text width=1.8cm, rounded corners=3pt, line width=0.3mm, fill=red!10, draw=red!80, font=\footnotesize},
    character_work/.style =    {align=center, text width=10cm, rounded corners=3pt, line width=0.3mm, fill=red!10, draw=red!0, font=\footnotesize},
    personalization/.style =           {align=center, text width=1.8cm, rounded corners=3pt, line width=0.3mm, fill=cyan!10, draw=cyan!80, font=\footnotesize},
    personalization_work/.style =      {align=center, text width=10cm, rounded corners=3pt, line width=0.3mm, fill=cyan!10, draw=cyan!0, font=\footnotesize},
    risk/.style =         {align=center, text width=1.8cm, rounded corners=3pt, line width=0.3mm, fill=orange!10, draw=orange!80, font=\footnotesize},
    risk_work/.style =    {align=center, text width=10cm, rounded corners=3pt, line width=0.3mm, fill=orange!10, draw=orange!0, font=\footnotesize},
}

\usepackage{CJK}

\definecolor{lightgreen}{RGB}{144,238,144}
\definecolor{darkcyan}{HTML}{008B8B}

\newcommand{\sys}{\textsc{AReaL2.0}\xspace}

\title{Next-Generation Agentic Reinforcement Learning Systems Enable Self-Evolving Agents}

\author{Ran Yan$^{1,2}$, Wei Fu$^{1,3}$, Jiale Li$^1$, Shusheng Xu$^1$, Zhiyu Mei$^1$, Jiaxuan Gao$^{1,3}$, Jiarui Zhang$^{1,2,3}$, Wentai Zhang$^1$,
Hao Dai$^1$,
Xujie Shen$^1$, Chuyi He$^1$, Zhen Pu$^1$, Jun Mei$^1$, Zhiyao Lin$^1$, Haitao Wang$^1$, 
Zhiqiang Ding$^1$, Jiawei Zhang$^1$, Huaijie Wang$^{1,3}$, Ruida Xu$^1$, Honghua Dong$^1$, Youhe Jiang$^2$, 
Yi Wu$^3$, Tongkai Yang$^1$, Binhang Yuan$^{1,2}$}

\affiliation{$^1$Ant Group, $^2$HKUST, $^3$Tsinghua University}
\vspace{-1.0em}

\abstract{
Large language model (LLM) agents are rapidly being deployed in production, including coding assistants, customer-support chatbots, and scientific research assistants, yet they remain fundamentally static in large-scale enterprise-level deployment. The LLM weights, system prompts, tool repertoires, and in-context harnesses are frozen at deployment time, and any improvement requires a manual loop of human-curated data collection, offline fine-tuning, modification of the agentic paradigm, and re-deployment. Recent work on self-evolving agents, such as OpenClaw for individual users, indicates that the next leap in agent capability will come from agents that continually learn from their own experience. In this paper, we argue that this vision for self-evolving agent deployment is being held back for enterprise-level large-scale agentic service not by reinforcement learning (RL) algorithms but by agentic online RL systems. Specifically, we argue that current agentic RL systems and the surrounding observability software stack are inadequate along three essential aspects: (\underline{\textbf{i}}) there is no standardized \textit{agent trajectory data protocol} capable of carrying RL learning signals at step granularity across heterogeneous agent paradigms; (\underline{\textbf{ii}}) there is no enterprise-grade comprehensive \textit{data proxy} that converts real workloads into governed learning substrates; and (\underline{\textbf{iii}}) there is no \textit{unified agent evolution control plane} with automatic triggering that decides, on the basis of trajectory statistics, when to update policy model weights or evolve the in-context harness by the corresponding RL algorithm. We argue that the next generation of agentic RL systems must be co-designed around these three pillars, and we sketch concrete architectural commitments, case studies, and counter-arguments to substantiate our position. We further instantiate one scoped branch of our vision through \sys, showing how existing RL infrastructure can be reorganized into an agent service-oriented online RL loop for policy LLM weight updates from deployed online agent workloads.

\vspace{-2.0em}
}

\begin{document}
\maketitle

\begin{center}
\vspace{-1.5em}
\href{https://github.com/areal-project/AReaL}{%
  \faGithub\,\texttt{\:Code: https://github.com/areal-project/AReaL}}
\vspace{0.5em}
\end{center}

\section{Introduction}
\label{sec:intro}

Large language model (LLM) agents are moving from laboratory demonstrations into deployed systems for research assistance, software engineering, and task automation. In fact, such LLM agents should essentially change the unit of deployment for LLM services at enterprise level---the deployed object should no longer be only a LLM that maps a sequence of input tokens to a sequence of output tokens; instead it should increasingly be a long-horizon agent policy embedded in a heterogeneous complicated bussiness environment, where the agent reads files, calls tools, retrieves documents, invokes APIs, updates memory, requests human approval, and accomplishes complicated analytic jobs. This shift creates a \textit{mismatch between how agents are deployed and how agents are improved}: A deployed enterprise agent may perform thousands or millions of tool-mediated interactions, yet its future behavior usually changes only through a manual engineering cycle: operators inspect traces, define new evaluation benchmarks, edit prompts, add tool descriptions, update a LLM through reinforcement learning (RL), implement a new agentic loop, and redeploy. On the other hand, we have recently witnessed that self-evolving agents are no longer hypothetical---personal agent systems, such as OpenClaw~\cite{openclaw}, and learning frameworks such as OpenClaw-RL~\cite{wang2026openclaw}, suggest that the next leap in agent capability will come from agents that continually learn from situated experience~\cite{xia2026metaclaw,ma2026skillclaw}. 

We interpret a \textit{self-evolving agent} as a deployed agent whose future behavior can improve as a consequence of its prior situated experience. To be cautious, we do not mean unrestricted recursive self-modification; instead, we mean a bounded closed loop in which each user’s interaction trajectories with the agent can be observed, redacted, verified, attributed, and converted into one of several governed updates: a memory insertion, a skill patch, a harness edit, a tool-schema modification, or an on-policy RL update. Recent agents built from trajectory reflection~\cite{shinn2023reflexion}, memory extraction~\cite{zhou2026memento}, context/harness edit~\cite{zhang2026agentic}, skill libraries~\cite{ma2026skillclaw}, and self-feedback~\cite{wang2026openclaw} already demonstrate that execution traces can be turned into reusable guidance. 

However, current self-evolving agent examples remain closer to individual-assistant or algorithm-specific self-improvement than to general enterprise-scale self-evolving agent services. For example, OpenClaw~\cite{openclaw} is compelling because it reframes the assistant as a persistent, user-owned agent that can perform tasks across inboxes, calendars, flights, and messaging apps, but this is not the same as a governed enterprise learning substrate spanning many teams, tenants, workflows, and compliance boundaries. On the other hand, an enterprise coding agent should learn from issue-resolution traces across teams while respecting repository permissions, license boundaries, proprietary code restrictions, and audit requirements; a customer support agent should learn from escalations, refunds, satisfaction signals, and policy updates while avoiding the leakage of customer data; a scientific agent must learn from failed experiments and literature-search trajectories while preserving provenance, reproducibility, and laboratory constraints. We believe that at the enterprise level, the missing component is not a larger model, a cleverer prompt, or even a specific RL algorithm; it is an enterprise-level system substrate that supports the smooth transfer from agentic interaction trajectories to self-evolving online learning. The current research focuses on agents that often prioritize improvements to better models, better prompting, and better toolkit integration, but this framing misses the system challenges for real-world large-scale deployment. 

Self-evolving, agentic RL should therefore be explored not only as an algorithmic family but also as a system discipline for design and implementation: how to capture experience, transform it into credit-assignable data, select an evolution mechanism, and deploy the resulting update appropriately. In this paper, we argue that \textit{self-evolving enterprise agents require next-generation online agentic RL systems}, and that such systems must have three co-designed pillars, including (\underline{\textbf{i}}) a standardized, vendor-neutral, RL-grade agent trajectory data protocol capable of carrying step-level learning signals across heterogeneous agent paradigms; (\underline{\textbf{ii}}) an enterprise-grade data proxy that converts real agentic workloads into governed, replayable learning corpus; (\underline{\textbf{iii}}) an unified agent evolution control plane that automatically decides, from trajectory statistics and operational constraints, whether to update memory, patch a skill, edit a harness, or update model weights through post-training RL. Concretely, we enumerate these three pillars below:

\underline{\textbf{First}}, self-evolving agents require a standardized \textit{agent trajectory data protocol} (ATDP) that makes the deployed agent experience learnable rather than merely observable. Existing agent logs typically record prompts, completions, tool calls, latency, errors, and token usage; such traces are useful for debugging but insufficient for online agentic RL. A learning-ready trajectory must instead preserve the decision process at step granularity: the observation available to the agent, the relevant internal or harness state, the chosen action, the action outcome, late-arriving reward or critique signals, and metadata such as model version, tool schema, tenant, cost, and governance status. The purpose of ATDP is therefore to turn heterogeneous agent executions into typed, auditable, replayable, and credit-assignable event sequences. This protocol should be vendor-neutral and framework-agnostic, support delayed feedback and reward augmentation, preserve versioned provenance for counterfactual replay, and encode privacy, access-control, retention, and training-eligibility fields from the beginning. Without such a protocol, enterprise agents may generate vast amounts of interaction data, but the data will remain too incomplete, non-reproducible, or unsafe to serve as the substrate for self-evolution.

\underline{\textbf{Second}}, self-evolving agents require an enterprise-grade \textit{agentic data proxy} that captures ATDP trajectories from real production workloads across models, tools, memory systems, retrieval systems, and human-feedback channels. ATDP specifies what must be represented; the data proxy specifies how those representations are produced in heterogeneous enterprise environments. Because deployed agents will not be built on a single framework, LLM provider, tool stack, or orchestration paradigm, the data proxy must intercept at stable execution boundaries: LLM calls, tool invocations, retrieval calls, memory reads and writes, file or browser actions, approval events, user corrections, and final task outcomes. Its role is not simply to export traces but to convert production work into governed learning material by redacting sensitive fields, enforcing access control and retention policies, attaching provenance, harvesting weak and delayed reward signals, and persisting trajectories in a form suitable for replay and training. Crucially, the data proxy must distinguish deterministic replay, approximate replay, and non-replayable events, because an enterprise learning system must be able to ask not only what happened, but whether the agent would have succeeded under a different prompt, model checkpoint, memory item, retrieval policy, or tool schema. In this sense, the data proxy serves as the first operational safety boundary of the self-evolving RL loop.

\underline{\textbf{Third}}, self-evolving agents require a \textit{unified agent evolution control plane} that decides when, where, and how agent behavior should change. Self-evolution should not be equated with blindly updating model weights. A deployed agent is a composite policy consisting of a base LLM, an in-context harness, memory, tools, and guardrails; different failures require different intervention surfaces. Recurring missing facts may call for memory insertion, tool-routing failures may call for harness or schema edits, reusable procedural failures may call for skill patches, and broad failures that persist across tenants, tasks, and tool configurations may justify model-weight updates through supervised fine-tuning, preference optimization, on-policy RL, process-reward learning, or distillation. The control plane should therefore treat self-evolution as a governed decision problem: given a window of ATDP trajectories, trajectory statistics, evaluator scores, user-correction rates, tool-failure clusters, cost signals, safety constraints, and distribution-drift indicators, it selects among memory updates, skill updates, harness edits, tool-schema changes, policy updates, rollback, or no-op. Each selected intervention must pass through replay-first evaluation, offline regression tests, tenant-aware safety checks, and versioned rollback. The control plane is thus the mechanism that turns captured experience into staged, auditable, and automatically triggered agent improvement, making enterprise-scale self-evolving agents possible.

Lastly, to ground this vision in a concrete system prototype, we instantiate one deliberately scoped branch of the proposed architecture in \sys. Rather than claiming to realize the full self-evolving agent substrate, \sys focuses on online policy LLM weight updates as a representative evolution path. This prototype shows how an existing RL framework (i.e., \textsc{AReaL}~\cite{fu2025areal}) can be reorganized from an offline post-training system into an agent service-oriented online RL loop: deployed agent services can redirect their LLM inference calls to \sys-managed agent-compute workers, while the resulting interaction trajectories are captured and consumed by the RL training pipeline. This design choice lets us explore the practical integration between agent serving, trajectory collection, and policy optimization, while leaving the broader multi-surface evolution problem---including memory updates, skill patches, harness edits, tool-schema evolution, replay-based governance, and automatic intervention selection---as the complete substrate system developing agenda.

\section{Related Work}

\textbf{RL for LLM post-training.} Reinforcement learning has been the state-of-the-art for LLM post-training~\cite{zhang2025survey}. RLHF demonstrated that fine-tuning with human preferences can improve instruction following~\cite{ouyang2022training}. Constitutional AI and RLAIF showed that AI-generated critiques and preference signals can substitute for some forms of human labeling~\cite{bai2022constitutional}. PPO remains a foundational policy-gradient method for constrained policy updates~\cite{schulman2017proximal}, while DPO reframes preference optimization as a simpler classification-style objective that avoids explicit reward-model training and online RL sampling~\cite{rafailov2023direct}. More recently, reasoning-oriented RL has renewed interest in online and outcome-driven optimization. DeepSeek-R1~\cite{guo2025deepseek} reported that reasoning behavior can be incentivized through RL, including emergent self-reflection and strategy adaptation.

\textbf{Self-evolving agents.} Self-evolving agents have emerged as a new paradigm for agentic workflows~\cite{zhang2025landscape}. ReAct~\cite{yao2022react} showed that LLMs can interleave reasoning traces and actions, giving rise to trajectories that are interpretable and tool-mediated rather than single-turn completions. Reflexion~\cite{shinn2023reflexion} converted task feedback into linguistic reflections stored in memory.  Memento-Skills~\cite{zhou2026memento} makes skills first-class evolving memory artifacts and updates them through a read-write reflective learning process. OpenClaw-RL~\cite{wang2026openclaw} focuses on personalized online signals from user interactions. 
These paradigms demonstrate the plausibility of learning from experience without always updating policy LLM weights as a new online agentic RL paradigm. 
The field, therefore, already has many mechanisms for ``what might evolve'': memory~\cite{zhou2026memento}, skills~\cite{ma2026skillclaw}, prompts~\cite{yuksekgonul2024textgrad}, search tool interfaces~\cite{gao2026unlocking}, and model weights. What the current landscape misses is an online RL system substrate for deciding, from deployed trajectories, which mechanism should evolve and how the resulting update should be evaluated, governed, and deployed.

\textbf{RL systems.} Recent work has optimized systems to improve the efficiency of RL-based post-training for LLMs~\cite{mei2024real,zhong2025optimizing,wu2025g}. HybridFlow/verl~\cite{sheng2025hybridflow} improves RL training throughput through a hybrid-controller execution model, a 3D-HybridEngine for zero-redundancy actor-model resharding, and automatic GPU placement. StreamRL~\cite{zhong2025streamrl} targets disaggregated RL training by mitigating pipeline bubbles caused by stage dependencies and skewness bubbles caused by long-tail generation lengths, while enabling heterogeneous and elastic resource allocation. AsyncFlow~\cite{han2025asyncflow} introduces asynchronous streaming dataflow and delayed parameter synchronization between rollout generation and policy updates, reducing hardware idling under bounded staleness. AReaL~\cite{fu2025areal} further decouples rollout generation from policy optimization through fully asynchronous execution, using staleness-aware training and a decoupled RL objective to maintain training stability. However, these systems are not sufficient to support self-evolving agents via online reinforcement learning. 

\textbf{Agent and RL data protocol.} Standardization has recently become an explicit research axis for both agent systems and RL datasets. Agent interoperability protocols such as MCP~\cite{mcp2024} and A2A~\cite{a2a2025} standardize how agents discover tools, access external context, and communicate across heterogeneous providers, while recent surveys classify these protocols primarily around context access, inter-agent coordination, security, scalability, and latency \cite{yang2025survey}. In parallel, RL data infrastructures such as D4RL~\cite{fu2020d4rl} and RLDS~\cite{ramos2021rlds} have shown the importance of standardized trajectory datasets for offline RL, imitation learning, replay, annotation, and reproducible benchmarking. Most directly related, Agent Data Protocol (ADP) proposes a lightweight “interlingua” for unifying heterogeneous LLM-agent datasets across tool use, browsing, coding, and general agentic workflows, demonstrating that standardized agent trajectories can improve scalable supervised fine-tuning~\cite{song2025agent}. However, these efforts do not target the problem addressed in this paper: deployed self-evolving agents require not only interoperability or offline dataset conversion, but an RL trajectory protocol that preserves step-level causal context, delayed reward signals, action outcomes, harness and tool versions, governance metadata, replay boundaries, and learning eligibility under enterprise constraints.

\section{Agent Trajectory Data Protocol}

In order to enable self-evolving agents, the first question to ask is:

\begin{quotation}
    \textit{What should be the right data abstraction that will be captured for learning?} 
\end{quotation}

To answer this question, our first step is to build a standardized agent trajectory data protocol (ATDP) capable of carrying RL-step-grade signals at step granularity across heterogeneous agentic workflows. We make the following key statement about this pillar of self-evolving agents: 

\begin{tcolorbox}[colback=blue!5!white,colframe=blue!75!black]
  \textbf{Key claim (\underline{i})}: \textit{Self-evolving agents require a standardized, vendor-neutral, RL-step-grade trajectory data protocol that captures the full lifecycle of agent decision process, including elements that current observability schemas usually discard.}

\end{tcolorbox}

\textbf{A prototype formulation.} A trajectory noted by $\bm{\tau}$ under ATDP is a typed event sequence defined as:
$$\bm{\tau} = (\mathbf{e}_1, \mathbf{e}_2, \ldots, \mathbf{e}_T).$$
And any intermediate step $\mathbf{e}_t$ is defined as:
$$\qquad \mathbf{e}_t = \langle \mathbf{o}_t, \mathbf{h}_t, \mathbf{a}_t, \mathbf{y}_t, r_t, \mathbf{m}_t \rangle$$ 
Where $\mathbf{o}_t$ is the \textit{observable status} (e.g., tool outputs, retrieval snippets, user message, environment state); $\mathbf{h}_t$ is the \textit{hidden internal status} (e.g., followed plan, scratchpad, confidence, reasoning summary); $\mathbf{a}_t$ is the chosen \textit{action} (e.g., tool call with typed argument schema, message in terms of generated tokens, code edit, memory update); $\mathbf{y}_t$ is the \textit{action outcome} for $\mathbf{a}_t$ (e.g., tool return, user accept/edit/retry/delete, or exit code);  $r_t$ is the reward signal (e.g., binary outcome, scalar score, natural language critique, or implicit signal extracted from $y_t$); $\mathbf{m}_t$ is the relevant \textit{metadata} (e.g., latency, tokens, cost, tenant, session, harness fingerprint, model id). Note that the schema reduces to a standard partially observable Markov decision process abstraction when $h_t = \emptyset$ and $m_t$ is dropped, but is rich enough to accommodate LLM-specific artifacts: reasoning traces, retrieval snippets, tool schemas, human corrections, natural language critiques, and rejected-action classes.

\textbf{Design principle of ATDP.} We enumerate the design principles for ATDP below:  

\begin{itemize}[leftmargin=*]
\item \textit{Decision-relevant bounded revelation}: ATDP should record enough information to improve behavior but should not require exposing every internal token or hidden reasoning trace. Concretely, a self-evolving agent needs access to enough structure to support credit assignment, causal diagnosis, and replay, but it does not generally need unrestricted disclosure of every hidden token or chain-of-thought fragment, which usually cannot provide auditable learning signals. 

\item \textit{Unification across frameworks and tasks}: Most existing agent logs are framework-specific, and most RL datasets are task-specific. Furthermore, most enterprise telemetry is optimized for debugging, reliability, and compliance, not for online agentic RL. One core innovation of ATDP is to make the unit of learning neither a prompt-response pair nor an opaque trace, but a typed, auditable, credit-assignable event record.

\item \textit{Credit assignability}: ATDP must make the trajectory possible to answer the question of ``which observation, prompt fragment, retrieval result, tool call, memory item, or guardrail decision contributed to success or failure?'',  which requires storing not only chosen actions but also the key decision context in which they were chosen. For example, for tool calls, ATDP should store tool version, argument schema, permission scope, latency, error class, return object, and whether the result was later trusted, ignored, corrected, or contradicted.

\item \textit{Late-bound learning signals}: Many useful rewards and critiques arrive after the acting step, e.g., a user correction in the next conversational turn, a failing test, a later human annotation, or a slower remote evaluator. ATDP should therefore allow an event’s reward field to be updated or augmented after initial logging while preserving immutability of the original causal record. This is essential for enterprise settings in which some judgment signals are asynchronous, policy-mediated, or sampled. ATEP should treat this as a first-class property of agent learning data. 

\item \textit{Versioned replayability}: Every event should be attributable not just to a model identifier, but to the exact execution envelope that produced it, including harness schema, tool version, retrieval index snapshot, guardrail configuration, and policy LLM version or checkpoint. Without these fields, ``agent experience'' becomes statistically useful but operationally non-reproducible. 

\item \textit{Governed observability}: Enterprise agentic interaction data could include privacy, security, and lineage fields from the beginning. ATDP should support redaction status, data-classification labels, tenant identifiers, retention policy, consent or legal basis, human-review status, and training eligibility. It should also support split visibility: a production debugger may see redacted traces, while a controlled training job may access sealed, policy-approved fields.

\end{itemize}

\section{Comprehensive Agentic Data Proxy}
\vspace{-0.25em}

Given the proposal of ATDP, the next question to ask is:

\begin{quotation}
    \textit{How to capture such data for enterprise-level heterogeneous agentic workflows?} 
\end{quotation}

Our answer is to implement a comprehensive data proxy. Concretely, we interpret the data proxy not merely as an API gateway, a tracing exporter, or a logging service; rather, it is the essential mechanism that converts production workloads into governed learning material. We introduce the second pillar of self-evolving agents as: 

\begin{tcolorbox}[colback=blue!5!white,colframe=blue!75!black]
  \textbf{Key claim (\underline{ii})}: \textit{Self-evolving agents require an enterprise-grade data proxy that can intercept, capture, anonymize, persist, and replay agentic workloads across heterogeneous frameworks and model providers, with lossless serialization for multiple agentic RL paradigms.}

\end{tcolorbox}

ATDP specifies what a learning-ready trajectory should contain, while the enterprise agentic data proxy specifies \textit{how such trajectories are captured in production}. The proxy sits between the agent and the LLMs (either deployed internally or by an external provider), between the agent and its tools, between the agent and its short- and long-term memory systems, and between the agent and human feedback channels. The key purpose is to transform production work into governed learning data without forcing every agent team to rewrite its application within an enterprise.

\textbf{Design principle of data proxy.} We introduce our key design of the data proxy as follows:

\begin{itemize}[leftmargin=*]
\item \textit{Existing framework-agnostic interception}: Enterprises usually will not standardize on a single agent framework. For example, some teams may use LangChain~\cite{langchain2025} or LangGraph~\cite{langgraph2025}; some teams will use CrewAI~\cite{crewai2025}; some teams will use the OpenAI Agents SDK~\cite{openai_agents_sdk2025} or Claude Agent SDK~\cite{claude_agent_sdk2025}; some will use MCP-connected tools~\cite{mcp2024}; others will write custom orchestration code. The data proxy must therefore intercept at stable boundaries: model API calls, tool invocations, retrieval calls, memory operations, file-system or browser actions, human approval events, and final user feedback.

\item \textit{Lossless ATDP emission}: Every intercepted call should be converted into an ATDP event stream and persisted in a form suitable for sequential learning by the data proxy. For LLM calls, lossless means preserving all fields that are eligible and necessary for learning under policy, including prompt-template fingerprint, system-prompt version, exposed tools, decoding parameters, sampled output, token IDs where available, log probabilities where available, and model version or checkpoints. For tool calls, Lossless means storing inputs, outputs, schema, version, permission scope, error class, and downstream use. 

\item \textit{Replay capability}: A trajectory that cannot be replayed is not trustworthy training data. The replay capability requires capturing tool inputs and outputs, environment state, LLM versions where licensable, file or database snapshots where permitted, and external side-effect boundaries. The replay capability also requires distinguishing deterministic replay, approximate replay, and non-replayable events. For example, A monitoring trace can include statements, such as ``The agent called tool X and failed.'' A training proxy must enable the system to ask: ``Would the agent have succeeded under a different prompt, model, memory, retrieval policy, or tool schema?'' Replay is the key feature that distinguishes a learning proxy from a monitoring proxy.

\item \textit{Cross-tenant aggregation with isolation}: An enterprise may want a single coding agent to improve from trajectories across multiple product teams, but each team may have different repository permissions, data-classification rules, and legal restrictions. The data proxy should therefore support isolated tenant stores, policy-based aggregation, federated or split-learning-style training jobs where needed, and tenant-aware evaluation. This is not only a privacy requirement but also a learning requirement: without tenant metadata and isolation, the system cannot know whether an update learned from one team should generalize to another.

\item \textit{Reward harvesting}: Agentic workloads contain many weak and delayed rewards: user replies, ticket reopen rates, test failures, compiler errors, human corrections, escalation decisions, refund reversals, approval latency, downstream edits, and abandonment. Following the central observation from OpenClaw-RL~\cite{wang2026openclaw} that every interaction can generate a next-state signal, such as a user reply, tool output, terminal state change, or GUI state change, an enterprise data proxy should generalize this observation by treating operational state changes as candidate learning signals. Not all signals are rewards, and not all rewards are safe to optimize directly, but the data proxy must capture them before the control plane can decide how to use them.

\item \textit{Data integrity before learning}. The data proxy should enforce redaction, access control, retention, and learning eligibility before the trajectory data enters its learning queues. This is the inverse of the common pattern in which logs accumulate first, and governance is attempted later. In a self-evolving system, the proxy is not merely an observability collector; it is the first safety boundary of the RL learning loop.

\end{itemize}

\section{Agent Evolution Control Plane}

Given the support of a comprehensive agentic data proxy, the last key question is about:

\begin{quotation}
    \textit{When and how to trigger corresponding RL optimization for self-evolving?} 
\end{quotation}

To answer this question, we propose a unified agent evolution control plane that automatically selects the appropriate intervention surface---memory, skill, harness, or weights---based on trajectory statistics, performance drift, risk, and cost. These are not auxiliary details; they are the system's preconditions for self-evolving agents at scale. In other words, self-evolution is not a single optimizer; rather, it is a decision problem under governance with multiple intervention surfaces. We summarize the third pillar of self-evolving agents as:

\begin{tcolorbox}[colback=blue!5!white,colframe=blue!75!black]
  \textbf{Key claim (\underline{iii})}: \textit{Self-evolving agents require a unified evolution mechanism that supports both model-weight updates through diverse RL algorithms and in-context harness engineering, and that automatically triggers the appropriate intervention based on trajectory statistics, performance drift, safety constraints, and cost.}

\end{tcolorbox}

\textbf{A prototype formulation.} We introduce a prototype formulation of the control plane. Formally, we define a deployed self-evolving agent $\mathcal{A}$ at time $t$ as:
$$
\mathcal{A}_t = \left\langle \bm{\pi}_{\theta_t},
\bm{\mathcal{H}}_{\psi_t},
\bm{\mathcal{M}}_t,
\bm{\mathcal{T}}_t,
\bm{\mathcal{G}}_t
\right\rangle .
$$
Here, $\bm{\pi}_{\theta_t}$ is the policy LLM parameterized by $\theta$, $\bm{\mathcal{H}}_{\psi_t}$ is the in-context harness (the harness policy parameterized by $\psi$), $\bm{\mathcal{M}}_t$ is memory, $\bm{\mathcal{T}}_t$ is the tool
repertoire and tool schemas, and $\bm{\mathcal{G}}_t$ is the safe governance and guardrail configuration. The targeted objects of self-evolution could be diverse, e.g, including system prompt instructions, developer instructions, agentic prompt templates, tool descriptions, routers, memory retrieval policies, memory policies, planning templates, subagent definitions, skill libraries, etc.

During self-evolving, the control plane observes a window of ATDP trajectories
$\bm{\mathcal{D}}_t = \{\bm{\tau}_i\}_{i=t{-}W}^{t}$ and selects an evolution action:
$$
u^\star =
\arg\max_{u \in \bm{\mathcal{U}}}
\left[
\mathcal{J}_{\mathcal{A}}(u \mid \mathcal{A}_t, \bm{\mathcal{D}}_t)
\right].
$$
Where $\mathcal{J}_{\mathcal{A}}$ generally measures the improvement of the evolution of the agent $\mathcal{A}$. And the evolving action set $\bm{\mathcal{U}}$ could include: 
(\underline{\textbf{i}}) update of policy LLM $\bm{\pi}_{\theta_t}$ (e.g., supervised fine-tuning, direct preference optimization, on-policy RL);
(\underline{\textbf{ii}}) update of in-context harness $\bm{\mathcal{H}}_{\psi_t}$ (e.g., skill patch, prompt edit);
(\underline{\textbf{iii}}) update of memory $\bm{\mathcal{M}}_t$ (e.g., retrieval-policy update);
(\underline{\textbf{iv}}) update of repertoire and tool schemas $\bm{\mathcal{T}}_t$ (e.g., tool-description edit, tool-schema change);
(\underline{\textbf{v}}) \textit{rollback}, and (\underline{\textbf{vi}}) no-op, under the safe governace control $\bm{\mathcal{G}}_t$.

\textbf{Control plane implementation considerations.} We summarize some key implementation considerations of the agent evolution control plan below:

\begin{itemize}[leftmargin=*]
\item \textit{Multi-surface adaptation}: The control plane should explicitly encode that different failure modes belong to different intervention classes. For example, if an agent's trajectories show recurring missing facts or reusable procedural lessons with a narrow scope, a memory insertion is usually the cheapest and safest update. If failures cluster around tool routing, retrieval formatting, guardrail wording, or developer-message structure, then harness editing is usually more appropriate; If, by contrast, the same category of failure persists across many tenants, tasks, and tool configurations, then the evidence suggests that the problem is probably not local to memory or harness and demands a policy update using RL, process-reward learning, or distillation.

\item \textit{Automatic triggering from trajectory statistics rather than anecdotal inspection}: The control plane should operate on explicit statistics such as evaluator scores, user correction rates, process-reward estimates, tool-specific failure clusters, canary deltas, cost per successful task, and drift in workload composition. We believe these signals would be useful, e.g., OpenClaw-RL turns next-state signals into both scalar process rewards and directional hints~\cite{wang2026openclaw}; AgentPRM shows the value of turn-wise rewards~\cite{xi2026agentprm}; and RLAnything combines step-wise and outcome signals~\cite{wang2026rlanything}. Thus, our essential technical consideration is to promote these statistics from monitoring artifacts into intervention triggers so that the self-evolution can be enabled by real-world agentic workflows.

\item \textit{Algorithm pluralism under a unified trigger interface}: We believe the next-generation agentic RL system should not hard-code a single particular learning paradigm---Concretely, for policy model weight update, the backend may involve on-policy or near-on-policy RL, process-reward learning, or on-police distillation from different directive feedback; for harness-space adaptation, the agentic RL system may involve prompt optimisation, context evolution, code search, or trajectory-aware verbal editing; for memory or skill updates, the agentic RL system may involve trajectory distillation into reusable workflow or reasoning objects. The control plane should organize such diverse paradigms instead of erasing these differences --- it decides which RL algorithm should be invoked, with which data slice, under which constraints, and on what deployment path.

\item \textit{Safe audited staged deployment}: Automatic triggering of agent self-evolution should not imply unreviewed hot-swapping of agent behavior. Every intervention should carry its own promotion path: shadow evaluation, retrospective replay checks, offline regression tests, canary rollout, rollback semantics, and differential monitoring against the incumbent version. This is especially important because the safety literature on personal agents already shows that persistent state, capabilities, and knowledge can create substantial attack surfaces. The right practice is not abandoning self-evolution, but ensuring evolution is stage-gated and auditable. A unified control plane is precisely the place where such gates belong.

\item \textit{Replay-first evaluation}: Before an agentic workflow update reaches users, the evolved agent should be evaluated against replayed trajectories and counterfactual variants. For example, a harness edit should be tested on past failures and known successes. A tool-schema change should be tested on tool-call replay. A policy model update should be tested on held-out trajectories, safety sets, tenant-specific evals, and distribution-shift probes. Note that this control-plane implementation also underscores why the data proxy’s replay capability is essential.

\item \textit{Versioned provenance and rollback}: We force every evolved artifact to be versioned: LLM checkpoint, prompt, tool schema, memory item, retrieval policy, guardrail, skill file, router, and dataset. Rollback should not be a manual emergency procedure; it should be a first-class action in evolving action set \bm{\mathcal{U}}. A self-evolving agent that cannot explain what changed is not self-evolving at the enterprise level---it is merely drifting.

\end{itemize}

\section{\sys: Prototype System Design and Implementation}

To make the preceding architectural claims more concrete \textit{without} claiming to realize the full landscope of the self-evolvoing agents substrate, we implement \sys as an exploratory prototype focused on one specific evolution path: online policy LLM weight updates from deployed agent trajectories. Our goal is not to implement the entire self-evolving agent substrate described above, nor to cover all possible intervention surfaces such as memory insertion, skill patching, harness editing, or tool-schema evolution. Instead, \sys uses policy LLM updating as a representative example to show how an existing RL framework can be modified from an offline post-training system into a simple online learning paradigm. Concretely, \sys exposes the rollout and training workers of the original \textsc{AReaL} framework as agent service-oriented computation components, allowing the existing agent service(s) to replace its standard LLM inference backend with a \sys-managed agent-compute worker. This design lets deployed agent interactions be served, captured, stored, and consumed by an online RL training loop with minimal changes to the surrounding agent harness. 

\textbf{\sys core design.} The core design goal of \sys is to adapt the rollout and training workers in current RL systems as a replacable inference service backbone into any deployed online agent-service without modifying any existing agent harness implementation. In other word, we follow the core desgin:

\begin{tcolorbox}[colback=green!5!white,colframe=green!75!black]
  \textbf{\sys prototype core desgin}: \textit{Instead of treating RL infrastructure as an offline post-training pipeline that is disconnected from online deployed agent workloads, \sys exposes the same computation units (e.g., GPU workers) used for rollout and policy training as micro-service components that can be plugged into any existing online agent deployment.}

\end{tcolorbox}

Concretely, \sys introduces an elegant lightweight extension to the current \textsc{AReaL} framework, where the original rollout and training workers are wrapped behind an agentic micro-service abstraction, allowing any existing agent service to replace its standard LLM inference backend (e.g., local deployment) with an \sys-managed agent-compute worker. This enables agentic trajectories generated by deployed agents interactions to be captured, stored, and consumed by the online RL training loop without requiring the agent application itself to be rewritten around a different offline RL runtime.

\begin{figure}[t!]
  \centering
  \includegraphics[width=\linewidth]{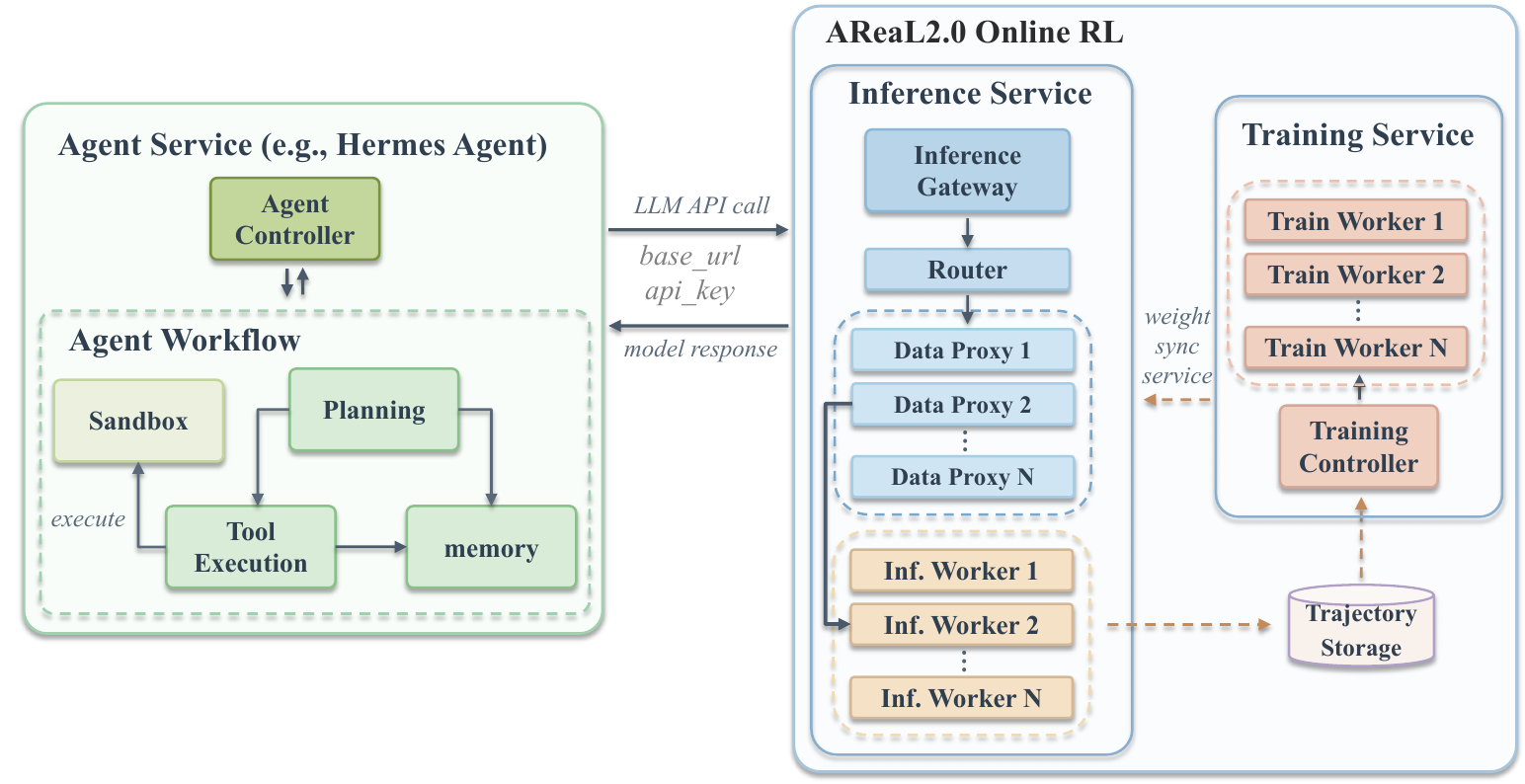}
  \caption{\textbf{\sys online RL workflow illustration}: An existing agent service keeps its original planning, tool execution, sandbox, and memory modules, while redirecting LLM API calls to the gateway, then the router routes requests through data proxies to agent-compute workers, records deployed trajectories, and connects inference serving with online RL training for policy-model weight updates.}
  \label{fig:areal2}
  
\end{figure}

\textbf{\sys implementation.} Our prototype system realizes this abstraction through four core components: a \textit{gateway}, a \textit{router}, a \textit{data proxy}, and an \textit{agent-compute worker}. Together, these components decouple public service protocols, session placement, trajectory management, and model inference or training computation (See Figure~\ref{fig:areal2}). This separation is essential for agentic RL workloads, where requests are often multi-turn, tool-augmented, and stateful, while training requires structured trajectories with token-level metadata and reward signals.

\begin{itemize}[leftmargin=*]

\item \textit{Gateway}. The gateway is the public entry point of the agent-service stack in the online RL system. The core functionaility of the gateway is to expose \sys as a replacement of the standard inference backend to an existing deployable agent service. From the perspective of the application, replacing an ordinary inference endpoint, e.g., an SGLang or vLLM server, with the \sys gateway is sufficient to redirect LLM calls into the online RL-enabled runtime. Internally, the gateway normalizes external LLM requests, authenticates access, and forwards agent turns into the online RL service fabric. This gateway essentially minimizes integration cost: the agent service continues to communicate through regular inference-style APIs, while \sys gains visibility into the interaction stream needed for online RL.

\item \textit{Router}. The router provides lightweight session-affinity management for multiple online RL training jobs. Note that agentic RL workloads are naturally stateful: a single task may involve multiple turns, tool calls, intermediate observations, and delayed rewards. The router assigns each session to a data proxy and preserves this assignment across turns. This allows \sys to support many concurrent agentic RL workloads while maintaining consistency of session state. Concretely, the router manages placement and affinity metadata, leaving high-volume interaction traffic to the data proxy and worker layers.

\item \textit{Data Proxy}. The data proxy implements the control-plane management for agentic data traffic and storage. The data proxy mediates between external service requests and the compute worker, records agentic data trajectories (e.g., conversation histories, tool-call events and responses), and prepares trajectory data for downstream training. In this sense, the data proxy is the component that converts ordinary deployed-agent traffic into online RL-consumable experience. The data proxy ensures that agent interactions are not merely served to users but are also structured as training data with the information needed for policy improvement.

\item \textit{Agent-Compute Worker}. The agent-compute worker is the execution abstraction that connects deployed agent services to the rollout and training backends in \sys. The core function is to wrap the various inference rollout engines, such as SGLang and vLLM, together with training workers such as Megatron- or FSDP-based actors, behind a micro-service interface. This worker abstraction allows compute resources to be dynamically allocated and deallocated according to the trajectory stream and training demand. As a result, \sys can serve online agent requests, collect trajectories, and update the underlying policy within a unified service-oriented architecture.
\end{itemize}

\textbf{Motivating example: online RL for hermes agent by \sys}.
We implement a representative use case, i.e., supporting the online RL training for the Hermes agent service~\cite{nousresearch2026hermesagent}. In a conventional deployment, Hermes invokes an inference backend, e.g., an SGLang worker, to obtain model responses during agent execution. With \sys, this backend can be replaced by an \sys-managed agent-compute worker exposed through the gateway. The surrounding agent service remains mostly unchanged: the agent service continues to issue LLM inference requests as before, while \sys intercepts the interaction stream, records trajectories, and connects them to the online RL training loop.
This example illustrates the central benefit of the \sys design. Rather than building a separate RL environment that attempts to imitate production behavior, the system reuses data generated by the original agent service itself. The agent’s native workflow, including multi-turn interaction and tool use, becomes the source of agentic data protocol trajectories. This reduces the gap between offline post-training data and deployed agent behavior, and provides a practical path toward continuously improving agents from their own service traffic.



\textbf{Toward the compelete self-evolving agent system substrate}.
The current \sys prototype should be interpreted as a deliberately scoped proof of feasibility for online policy-model adaptation, rather than as a complete implementation of the self-evolving agent substrate visioned by the complete lanscope of the self-evoloving agent system substrate. We demonstrate one important integration principle: existing RL infrastructure can be lightly reorganized so that rollout generation, inference serving, trajectory collection, and policy optimization are connected to deployed agent workflow with minimal change of the originl online deployment. However, this prototype only covers the model-weight-update branch of the broader control-plane action space. A full self-evolving agent system still requires several capabilities beyond the current prototype, including a complete ATDP implementation with step-level decision context and governance metadata, a comprehensive data proxy that captures tool, retrieval, memory, file, browser, human-feedback, and delayed-reward events, replay and counterfactual evaluation support, tenant-aware privacy and training-eligibility enforcement, and an evolution control plane that can automatically choose among memory updates, skill patches, harness edits, tool-schema changes, policy updates, rollback, or no-op. Thus, \sys should be viewed as an initial system step that grounds the policy-update case, while the full multi-surface, governed, replayable, and automatically triggered self-evolution loop remains the larger research and systems agenda.

\section{Conclusion}

We argued that self-evolving agents can be enabled by the next generation of agentic RL systems. Our key position is that the foremost bottleneck for enterprise-scale self-evolving agents is not only the absence of more powerful LLMs or more effective RL algorithms, but the absence of a system substrate that can transform deployed agent experience into governed, credit-assignable, and replayable learning material. To make this possible, we proposed three co-designed pillars: a standardized agent trajectory data protocol, an enterprise-grade agentic data proxy, and a unified agent evolution control plane. The \textit{first pillar}, the agent trajectory data protocol (ATDP), reframes agent experience as typed, step-level, RL-grade event data rather than opaque logs or prompt-response pairs. Such a protocol should preserve the decision context, action, outcome, reward signal, metadata, provenance, and governance status needed for credit assignment and replay. The \textit{second pillar}, the comprehensive agentic data proxy, specifies how heterogeneous production interactions across models, tools, retrieval systems, memory stores, and human-feedback channels can be intercepted, redacted, persisted, annotated, and replayed as learning-ready trajectories. The \textit{third pillar}, the agent evolution control plane, determines when and how the agent should evolve by inserting memory, patching skills, editing a harness, changing a tool schema, updating policy LLM weights through RL, rolling back, or doing nothing. Together, these three pillars shift self-evolving agents from an algorithmic aspiration to an enterprise systems problem. As a concrete initial step, we implement \sys that demonstrates the policy model update branch of this agenda by connecting deployed online agent serving, trajectory collection, inference workers, and RL training workers within one online learning workflow---the full self-evolving agent substrate still requires complete ATDP support, governed replay, and automatic multi-surface evolution beyond policy-model weight updates.

\bibliographystyle{unsrt}
\bibliography{references}


\end{document}